\documentclass[twocolumn]{aastex701}

\usepackage{chngcntr}
\usepackage{subcaption}
\usepackage{graphicx}
\usepackage{multirow}
\usepackage{hyperref} 
\usepackage{lineno}
\linenumbers

\begin{document}

\title{Reanalysis of the eclipses of LHS\,1140\,c: No evidence of an atmosphere and implications for the internal structure of the planet}

\correspondingauthor{Alexandra Rochon}
\author[orcid=0009-0008-1229-3230, gname='Alexandra', sname='Rochon']{Alexandra Rochon}
\affiliation{Institut Trottier de recherche sur les exoplanètes, Université de Montréal, 1375 Ave Therèse-Lavoie-Roux, Montréal, QC, H2V 0B3, Canada}
\affiliation{Department of Physics \& Astronomy, McMaster University, 1280 Main St W, Hamilton, ON, L8S 4L8, Canada}
\email[show]{alexandra.rochon@mail.mcgill.ca} 

\author[0000-0003-3506-5667]{Etienne Artigau}
\affiliation{Institut Trottier de recherche sur les exoplanètes, Université de Montréal, 1375 Ave Therèse-Lavoie-Roux, Montréal, QC, H2V 0B3, Canada}
\affiliation{Observatoire du Mont-Mégantic, Université de Montréal, Montréal H3C 3J7, Canada}
\email{etienne.artigau@umontreal.ca}

\author[0000-0002-7992-469X]{Drew Weisserman}
\affiliation{Department of Physics \& Astronomy, McMaster University, 1280 Main St W, Hamilton, ON, L8S 4L8, Canada}
\email{weisserd@mcmaster.ca}

\author[0000-0003-4987-6591]{Lisa Dang}
\affiliation{Institut Trottier de recherche sur les exoplanètes, Université de Montréal, 1375 Ave Therèse-Lavoie-Roux, Montréal, QC, H2V 0B3, Canada}
\affiliation{Waterloo Centre for Astrophysics and Department of Physics and Astronomy, University of Waterloo, 200 University W, Waterloo, Ontario, Canada, N2L 3G1}
\email{lisa.dang@uwaterloo.ca}

\author[0000-0001-5485-4675]{René Doyon}
\affiliation{Institut Trottier de recherche sur les exoplanètes, Université de Montréal, 1375 Ave Therèse-Lavoie-Roux, Montréal, QC, H2V 0B3, Canada}
\affiliation{Observatoire du Mont-Mégantic, Université de Montréal, Montréal H3C 3J7, Canada}
\email{rene.doyon@umontreal.ca} 

\author[0000-0001-9291-5555]{Charles Cadieux}
\affiliation{Institut Trottier de recherche sur les exoplanètes, Université de Montréal, 1375 Ave Therèse-Lavoie-Roux, Montréal, QC, H2V 0B3, Canada}
\affiliation{Observatoire de Gen\`eve, D\'epartement d’Astronomie, Universit\'e de Gen\`eve, Chemin Pegasi 51, 1290 Versoix, Switzerland}
\email{charles.cadieux.1@umontreal.ca}

\author[0000-0001-5383-9393]{Ryan Cloutier}
\affiliation{Department of Physics \& Astronomy, McMaster University, 1280 Main St W, Hamilton, ON, L8S 4L8, Canada}
\email{ryan.cloutier@mcmaster.ca}


\begin{abstract}

We present the reanalysis of three 15\,$\mu$m JWST/MIRI secondary eclipses of LHS\,1140\,c, a warm super-Earth (R$_{\rm{p}}$\,=\,1.272\,R$_{\oplus}$) in a 3.78-day orbit around an M4.5 dwarf.
We present a novel method for data reduction that leverages spatial derivatives of the point-spread function and compare it to widely used aperture photometry. 
Both methods yield eclipse depth consistent within 1$\sigma$ of the values reported in the literature. 
We measure an eclipse depth of 271$^{+31}_{-30}$\,ppm corresponding to a brightness temperature of $T_B=595^{+33}_{-34}$\,K, consistent with a bare rock. The secondary eclipse occurs 4.1$\pm$0.8\,minutes before the circular-orbit predicted time.
We explore the implications of our results on the internal structure of LHS\,1140\,c, the orbital architecture of the system and the possibility of future observations with JWST. We find a core-mass fraction (CMF) informed by the stellar abundances of refractory elements of 0.34$\pm0.11$, inflated compared to the CMF from radius and mass measurements, suggesting the possible presence of bulk volatiles in the interior.


\end{abstract}

\section{Introduction} 

The James Webb Space Telescope (JWST) is revolutionizing our ability to put constraints on the presence of atmospheres around temperate, terrestrial planets orbiting M dwarfs \citep[e.g.,][]{cadieux2024b, Damiano2024, espinoza2025, Glidden2025}.
Indeed, the small radii, low masses, low luminosities and high abundance of M dwarfs provides a favourable opportunity for characterizing the planets that orbits them with methods such as transit, secondary eclipse, and radial velocity observations. However, M dwarfs are also highly active, especially in their first $\sim$ 1 Gyr, emitting extreme-ultraviolet (XUV) radiation and flares, which calls into question the survivability of atmospheres around rocky planets \citep{shields2016}.
%
Although unravelling a star's historic flaring rate and cumulative XUV radiation is challenging, it is crucial for understanding how this history has shaped the present-day atmosphere of the planets it hosts. Fortunately, characterizing multi-planet systems can provide valuable insights into their evolutionary past as they share a common stellar evolution. Therefore, they offer a unique opportunity to improve our understanding of atmospheric escape mechanism \citep{Krissansen-Totton2023}.
%

Orbiting a calm, old ($>$5 Gyr) M4.5V star, the LHS\,1140 system is a benchmark for M-dwarf planetary habitability. 
It is the second-closest known system that hosts a habitable zone planet after TRAPPIST-1, with two transiting planets. 
%
%
First discovered in 2017, the outer planet, LHS\,1140\,b, is a  super-Earth (M$_{\rm{p}}$\,=\,5.60$\pm$0.19\,M$_{\oplus}$, R$_{\rm{p}}$\,=\,1.730$\pm$0.025\,R$_{\oplus}$) with a period of 24.7\,days \citep{dittman2017, cadieux2024}. 
A second rocky planet, LHS\,1140\,c (M$_{\rm{p}}$\,=\,$1.91\pm0.06$\,M$_{\oplus}$, R$_{\rm{p}}$\,=\,1.272$\pm$0.026\,R$_{\oplus}$) was discovered with a shorter 3.8-day orbital period using Spitzer photometry \citep{Ment2019, cadieux2024}. 
\citet{lillo-box2020} revisited the system using ESPRESSO \citep{pepe2021} radial velocity and TESS transits to improve the mass constraints and orbital properties of the planets. They also find evidence for a third non-transiting planet in the system at a \hbox{78.9-day} orbital period.
The TESS observations, in combination with previous unpublished Spitzer observations (PI: J.A.Dittmann, PID: 13174), the double transit observed by \citet{Ment2019} and HST observations (PN: 14888; PI: J. A. Dittmann; \citeauthor{Edwards2021} \citeyear{Edwards2021}) were jointly analyzed by \citet{cadieux2024}, greatly improving radii constraints for both transiting planets. They also reanalyzed the ESPRESSO radial velocities to derive very precise density measurements. A double transit of planets b and c was observed with NIRISS/SOSS by \citet{cadieux2024b} ruling out hydrogen-rich atmosphere for LHS\,1140\,b, but finding tentative evidence of an N$_2$-rich atmosphere. 
NIRSpec transmission spectroscopy observations of LHS\,1140\,b are also inconsistent with a hydrogen-rich atmosphere \citep{Damiano2024}. 
Internal structure models of LHS\,1140\,b are consistent with two degenerate scenarios: a water world with a large water-mass fraction or an airless rocky world without a core \citep{cadieux2024, cadieux2024b}. 

The first secondary eclipse observations of LHS\,1140\,c presented in \citet{fortune2025} are consistent with no significant atmosphere on this planet.
LHS\,1140\,c is close to the traditional cosmic shoreline described by \citet{zanhle2017} and is expected to be a bare rock, i.e., without a significant atmosphere \citep{redfield2024}.
%
This paper presents a new method to analyze these same observations. The eclipses of LHS\,1140\,c were also part of a uniform reanalysis using Frame-Normalized Principal Component Analysis (FN-PCA), which supported the conclusion that the planet is a bare rock \citep{connors2025}.
However, LHS\,1140\,b's high surface gravity and low insolation place it well within the cosmic shoreline, making it a prime candidate for atmospheric retention. 
Extended modeling of the Cosmic Shoreline identifies LHS\,1140\,b as the highest priority target to retain a CO$_2$-dominated atmosphere \citep{ji2025}. \citet{Berta-Thompson2025} estimate a 99\% probability that LHS\,1140\,b has an atmosphere.
Studying both planets in this system can shed light on atmospheric retention mechanisms because of the shared stellar evolution \citep{Krissansen-Totton2023}. The absence of an atmosphere on planet c is an opportunity to study its interior without the complications that an atmosphere poses.
LHS\,1140\,b was also announced as a target of the Director's Discretionary Time (DDT) \textit{Rocky Worlds} program to be observed in the coming year at 15\,$\mu$m.\footnote{https://rockyworlds.stsci.edu} 

The method of time-series mid-infrared eclipse photometry is becoming a powerful tool to detect atmospheres around small, rocky planets \citep{redfield2024}.
Specifically, the technique of eclipse photometry assumes that the planet is tidally locked to its host star and observes the thermal energy emitted by the dayside of the planet during the secondary eclipse, as it passes behind the star. The measured decrease in brightness, expressed as the ratio of the measured planetary flux to stellar flux, $F_p$\,/\,$F_*$,  corresponds to the occulted dayside thermal emissions. This is directly related to the brightness temperature of the planet. 
This method is immune to stellar activity, such as flares and the transit light source effect (TLS; \citeauthor{Rackham2018} \citeyear{Rackham2018}), a limiting factor in the transit spectroscopy method \citep{lim2023, moran2023, cadieux2024, cadieux2024b}. 

The presence of an atmosphere allows for heat redistribution to the nightside, reducing the eclipse depth $F_p$\,/\,$F_*$ compared to the bare rock scenario \citep{mansfield2019, koll2019, koll2022}. JWST's (\citealt{gardner2006}) mid-infrared instrument \citep[MIRI;][]{Rieke2015}, is equipped with a 15\,$\mu$m filter that overlaps with a CO$_2$ broadband feature. 
Hence, the presence of CO$_2$ in the atmosphere of a planet would further decrease the planet's brightness at 15\,$\mu$m \citep{cadieux2024b}. 
%

MIRI time series observations are known to be affected by detector systematics characterized by a settling ramp at the beginning of the observations \citep[e.g.,][]{zieba2023, greene2023, august2025, Meier-Valdes2025, fortune2025} and does not affect all pixels uniformly \citep{fortune2025}.
This often requires the discarding of the first 30-60\,minutes from the analysis. Recent investigations found that the brightness of the target \citep{connors2025} and the filter used during the preceding observations \citep{fortune2025} are factors affecting the detector settling time.

As these systematics are often more important than the predicted eclipse depths, we need to ensure that the eclipse depth inferences are robust to a variety of methods to detrend instrumental noise. 
In this work, we present the re-analysis of the secondary eclipses of LHS\,1140\,c observed by \citet{fortune2025} with a novel, data-driven photometry extraction pipeline, the MIRI Analysis Module (\texttt{MIRIAM}). We also detrend the detector systematics using parametric models individually picked for each eclipse, to account for the variability in the systematics between observations.
Finally, we present insights into the orbital configuration of the LHS\,1140 system and the implication of our results on the internal structure of LHS\,1140\,c.




We present the observations in Section \ref{sec:observation}. We explain the two methods used to extract the light curves and how we model the eclipse and detector systematics in Section \ref{sec:method}. 
The detrended light curves are shown and the results of the two methods are compared in Section \ref{sec:results}. We discuss the implications of our results on the internal structure of LHS\,1140\,c, and on the orbital architecture of the system in Section \ref{sec:discussion}. Our conclusions are summarized in Section \ref{sec:conclusion}.






\section{Observations} \label{sec:observation}
LHS\,1140 was observed during the secondary eclipse of planet c using JWST/MIRI with the F1500W filter centered on 15\,$\mu$m.
The three eclipses of  were obtained on November 27$^{\rm th}$ 2023, July 7$^{\rm th}$ 2024 and July 19$^{\rm th}$ 2024 as part of the JWST Cycle 2 program, Hot Rocks Survey (GO: 3730 PI: Diamond-Lowe, Co-PI: Mendonça). Each visit lasted $\sim$233\,minutes (exposure time) and used the SUB256 array. It consisted of 36 groups per integration, for a total of 1262 integrations per exposure, at a cadence of 11.1\,s.
These observations were first analyzed in \citet{fortune2025}.

\begin{figure*}
\centering
\begin{subfigure}{1\textwidth}
    \includegraphics[width=\textwidth]{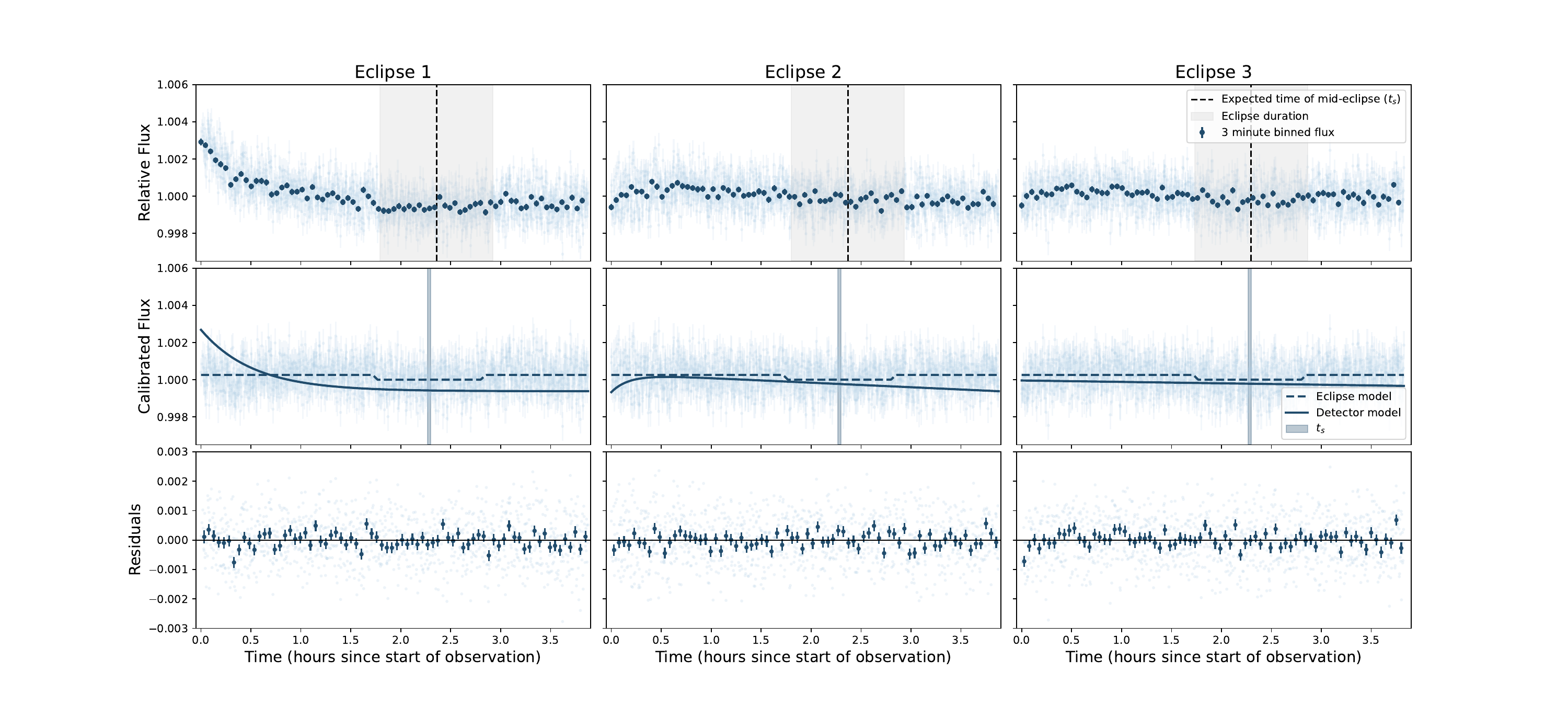}
    \caption{Best-fit model for the three light curves extracted using aperture photometry with \texttt{Eureka!}. Outliers beyond 4\,$\sigma$  are clipped.}
    \label{fig:raw-eureka}
\end{subfigure}
\hfill
\begin{subfigure}{1\textwidth}
    \includegraphics[width=\textwidth]{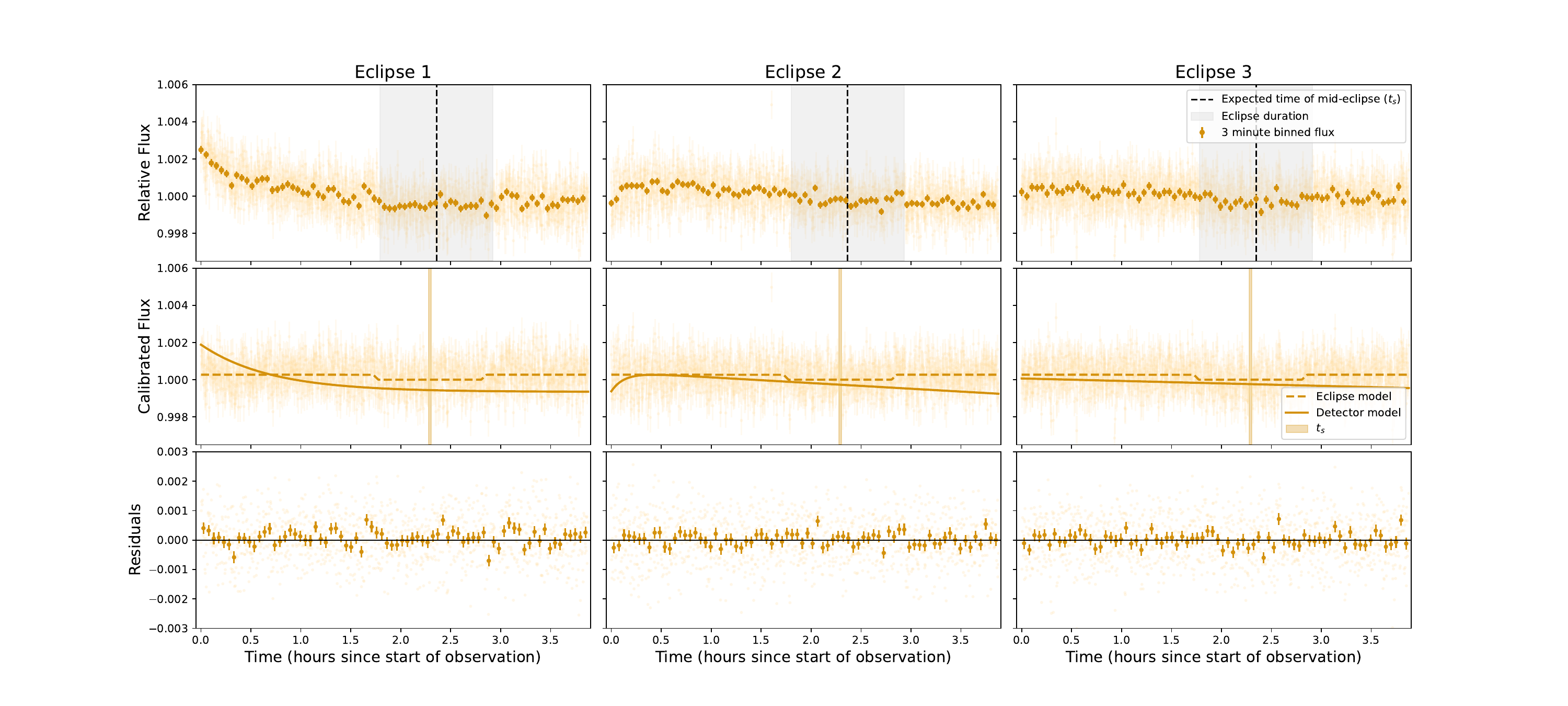}
    \caption{Best-fit model for the three light curves extracted using differential photometry with \texttt{MIRIAM}.}
    \label{fig:raw-miriam}
\end{subfigure}
    \caption{Comparison of the best-fit models for the three  eclipses (UT2023-11-27, left; UT2024-07-07, middle; UT2024-07-19, right) with the two light curve extraction methods (top and bottom).
    The raw photometry is plotted in light colour in the top panels with error. The error is calculated at the extraction step. Three-minute bins are shown in darker colour for clarity. The vertical black dashed lines indicate the expected time of mid-eclipse with the shaded grey region showing its duration.
    The corrected photometry with the best-fit detrending model for each eclipse is shown in the middle panels. The error is fitted with the MCMC. The dark, solid line corresponds to the detrending model and the dashed line to the jointly fitted astrophysical signal. The shaded light vertical band is the derived time-of-mid eclipse $t_s$ with uncertainty.
    The last panels shows residuals when the astrophysical model is subtracted from the corrected flux. Three-minute bins are overlain in darker color for clarity.
    The best fit systematics models are \texttt{E}, \texttt{DE}, \texttt{L}, for the three eclipses respectively, and are conserved for both extraction methods.
    }\label{fig:raw-eclipses}
\end{figure*}

\section{Data reduction}\label{sec:method}

We downloaded the uncalibrated `uncal' FITS files for the three eclipses from the Mikulski Archive for Space Telescopes (MAST) and can be found at \dataset[10.17909/vd64-n596]{http://dx.doi.org/10.17909/vd64-n596}.
We experiment with two different methods to extract the light curve, the commonly used aperture photometry described in Section \ref{sec:eureka} using the \texttt{Eureka!} pipeline\footnote{https://github.com/kevin218/Eureka.git} \citep{Bell2022}, and the new module \texttt{MIRIAM}, presented in Section \ref{sec:miriam}.
%
For both extraction methods, we fit the instrumental noise using detrending models outlined in Section \ref{sec:detrending_models} and we compare the results.

\subsection{Light Curve Extraction}\label{sec:extraction}
We first calibrate the raw data by running stages 1 and 2 of \texttt{Eureka!},
which are wrappers of initial steps of the \texttt{jwst} calibration pipeline\footnote{https://github.com/spacetelescope/jwst.git}. 
We use the default \texttt{jwst} settings, but we increase the jump detection threshold to 7$\sigma$ and skip the `photom' step to obtain better uncertainties at stage 3. We excluded the first and last groups of each integration, to fix a known issue with reset effect corrections outlined in \citet{morrison2023}. We experimented with keeping the first and last groups and determined that the residual root-mean-squared (RMS) of the individual fits were reduced by approximately 4-19\% when excluding them. 
We use the same `calints' FITS files (i.e., output of stage 2 described above) with both extraction methods.

\subsubsection{Aperture Photometry with \texttt{Eureka!}}\label{sec:eureka}
We perform aperture extraction on the `calints' FITS files during stage 3  of \texttt{Eureka!}. We compare a range of aperture sizes from 4 to 12 pixels and determine that an aperture size of 5 pixels reduces the scatter in the light curve and gives the best signal-to-noise ratio (SNR).
A background subtraction is performed with a fixed annulus centered at the point-spread-function (PSF) with inner radius of 12 pixels and outer radius of 30 pixels.
We mask pixels with an odd entry in the data quality (DQ) array.
We experiment with the stages 1 and 2 settings, following the methodologies of \citet{greene2023}, \citet{zieba2023} and \citet{fortune2025} respectively, and find that the latter reduces the RMS. The main differences are a jump detection threshold of 7$\sigma$ instead of 10$\sigma$ and removing the first and last group of each integration. We note that \citet{fortune2025} uses a larger radius for background subtraction. 
We clip 4$\sigma$ outliers at stage 3. 
%
The raw, aperture-extracted light curves for the three eclipses are shown in the top panels of Figure~\ref{fig:raw-eureka}. 
All three eclipses show a monotonic decreasing ramp over time with the first eclipse being more affected by this systematic. The shape of this ramp varies significantly between observations.

\subsubsection{Differential Photometry with \texttt{MIRIAM}}\label{sec:miriam}


The \texttt{MIRIAM} framework builds upon the \texttt{SOSSISSE} methodology \citep{lim2023, cadieux2024b}, originally developed for NIRISS/SOSS observations. In this framework, each integration is modeled as a linear combination of a reference PSF image and its spatial derivatives in the $x$ and $y$ directions. It also provides an explicit measurement of morphological changes in the PSF. More details about this method are provided in Appendix \ref{sec:miriam-explanation}.

\subsection{Detector Response Functions}
\label{sec:detrending_models}
We experiment with seven detrending models as a function of time, $t$: (1) a linear slope \texttt{L}, (2) a single exponential \texttt{E}, (3) a double exponential \texttt{DE}, (4) a third degree polynomial \texttt{P}, (5) a multiplicative combination of the linear slope and the single exponential \texttt{LE}, (6) a multiplicative combination of the linear slope and the third degree polynomial \texttt{LP}, and (7) a multiplicative combination of the single exponential and the third degree polynomial \texttt{EP}. The models, \texttt{L, E, DE} and \texttt{P} are described respectively as:
%
\begin{equation}
   L(t) =  a_1 \cdot t + a_2,
\end{equation}
%
\begin{equation}
   E(t) =  b_1 \cdot e^{(-b_2 \cdot t)} + b_3,
\end{equation}
%
\begin{equation}
   DE(t) =  (c_1 \cdot e^{(-c_2 \cdot t)} + c_3) + (c_4 \cdot e^{(-c_5 \cdot t)} + c_6),
\end{equation}
and
\begin{equation}
    P(t) =  d_1 \cdot t^3 + d_2 \cdot t^2 + d_3 \cdot t + d_4.
\end{equation}
$a_n, b_n, c_n$ and $d_n$ are fitted model coefficients. We also experiment with a second-order polynomial centroid model, independent of time, that relies on the change in the position of the PSF, denoted by $x_0$ and $y_0$: 
\begin{equation}
    D(x_0, y_0) =m_1 + m_2 x_0 + m_3 y_0 + m_4 x_0^2 + m_5 x_0 y_0 + m_6 y_0^2,
\end{equation}
where $m_n$ are fitted model coefficients. All combinations of detrending models are multiplied.

\subsection{Modeling the Time Series}
We fit each eclipse individually and perform an additional joint fit assuming the same time of mid-eclipse and eclipse depth.
We model the photometry as the product of the astrophysical signal and detector response, and simultaneously fit them using Markov Chain Monte Carlo (MCMC) sampling with \texttt{emcee} \citep{emcee}. The eclipse is fitted with \texttt{batman} \citep{Kreidberg2015} and the systematics are fitted using one of the detrending models described in Section \ref{sec:detrending_models}.

We initialize 70 walkers, 90\,000 burn-in steps and 100\,000 total steps.
We fix the ratio of planetary radius to stellar radius, $R_p/R_*$, the scaled semi-major axis, $a/R_*$, and the inclination, $i$, to the values reported by \citet{cadieux2024}.
%
The time of mid-transit, $t_0$, is fixed to the value of the ephemerides from the NASA Exoplanet Archive, calculated from \citet{cadieux2024}. 
We fix the eccentricity, $e$, to 0 and the argument of periastron, $\omega$, to 90, assuming a circular orbit. This is consistent with modeling suggesting a short circularization timescale for the LHS\,1140 planets \citep{gomes2020}. 

We fit for the eclipse depth ($F_p$) as the ratio of planetary flux to stellar flux, $F_p$\,/\,$F_*$, and for the time of mid-eclipse, $t_s$. 
We put an uninformative and uniform prior from $-1$ to 1 on $F_p$ for the joint fit, and 0 to 1 for the individual fits. Allowing for negative eclipse depth values prevents biasing towards an eclipse detection.
We put a uniform prior of half the duration of the eclipse on either side of the expected time of mid-eclipse for $t_s$.

We initialize the walkers to the value of 230\,ppm, which corresponds to the theoretical value of $F_p$ for a rocky, airless LHS\,1140\,c. $t_s$ is also initialized to the circular orbit prediction ($t_0 + P/2$).
We fit for the coefficients of the models detailed in Section \ref{sec:detrending_models}. The coefficients can take any value ($-\infty$, $+\infty$), but the walkers are initialized to a ``first guess" position estimated using \texttt{scipy.optimize.curve\_fit} \citep{Virtanen2020} or \texttt{numpy.polyfit} \citep{Harris2020} for the initial curve (out-of-expected-eclipse). The coefficients for the second-order polynomial centroid model are initialized to randomly selected values ([0.5, 0.1, 0.1, -0.1, 0.1, 0.1]) as they vary greatly depending on the combination of functions used to described the detector response.
%
%
We do not discard the first 30-60 minutes of observations as is customary with MIRI observations to let the ramp inform the detrending model. 
%
The code described here is available on GitHub\footnote{\url{https://github.com/arocho11/MIRI_Eclipse_Photometry_Analysis.git}}.

\subsubsection{Model Fitting and Comparison}

To ensure that the MCMC has converged, we verify that the distribution of the walkers for each parameter is approximately constant over the last 10\,000 steps. We do a visual inspection of the convergence. The uncertainty of the parameters are calculated from the 16$^{\rm th}$ and the 84$^{\rm th}$ percentiles of the converged steps.

To select the best-fit systematics model, we compare each model and their respective Bayesian information Criterion (BIC; \citeauthor{schwarz1978} \citeyear{schwarz1978}), defined as
\begin{equation}
    BIC = N_{par} \ln N_{data} -2 \ln L,
\end{equation}
where $N_{par}$ is the number of free parameters, $N_{data}$ is the number of data points, and $\ln L$ is the maximum log-likelihood from the MCMC posterior distribution defined as
\begin{equation}
    \ln(L) = -\frac{1}{2}\chi^2 - N_{data} \ln \sigma_F - \frac{N_{data}}{2} \ln(2\pi).
\end{equation}
A comprehensive model comparison is plotted in \hbox{Figure~\ref{fig:compare_plots}}, showing the corrected light curves for each detrending model, and the $\Delta$BIC calculated respective to the lowest BIC. 
Some models have too many degrees of freedom, so the parameters are not well constrained even when the MCMC runs for an extended number of steps. They produce a higher BIC and are discriminated against in the selection process.
We use a red-noise test to investigate the level of time-correlated noise in the residuals. We bin the residuals at increasing bin sizes and calculate the RMS of the binned residuals. We then verify that this is in good agreement with the expected residual RMS for pure white noise. We plot and compare the red-noise test for each model. A red-noise comparison figure can be found in Appendix \ref{sec:model_compare} for each eclipse and for both joint fits.
We also compare the RMS of each model and the errorbars and we run a $\chi^2$ test to assess the badness-of-fit.


\subsubsection{Individual Fits}\label{sec:individual}
We compare seven different detrending models. We also compare using no detrending with the PSF metrics or a second-order centroid model. We plot each model in the comparison figure found in Appendix \ref{sec:model_compare} for each eclipse. In the comparative figure, each column indicates the seven detrending models in time and each row corresponds to the detrending model, or lack thereof, in the centroid position, $x_0$ and $y_0$. We select the one which reduces the $\Delta$BIC. The best-fit model is boxed in dark blue and the color gradient indicates the fit preference (darker is better). We also compare the red-noise test for each model in Figure~\ref{fig:rednoise}. The black line in each plot represents the reduction in RMS scatter expected for purely white nose, whereas the coloured line is the RMS of the binned residuals at increasing bin sizes


\subsubsection{Joint Fit}

The eclipses are fit jointly using \texttt{batman}. The three eclipse therefore share an eclipse depth and time of mid-eclipse. Because \texttt{batman} is periodic, it can run on a single array that contains the three eclipses with their time series. 
While the astrophysical signal is fit jointly, the detrending models are uncorrelated between each visit and are applied independently to each eclipse with their own set of parameters. Since the systematics are visually distinct from one eclipse to the other, a single model could not fit each of them well. The detrending model used for each eclipse is the best individual fit, selected by picking the lowest BIC from the individual fits. 
%
%
The coefficients for these models are initialized to the preferred solutions of the individual fit before the \texttt{emcee} run.

\section{Results}\label{sec:results}

\begin{table*}
    \centering
    \caption{Comparison of the individually and jointly fitted eclipses from the two different extraction methods.}
    \label{tab:table1}
    \begin{tabular}{c|ccc|ccc}
        \hline
        \hline
         & \multicolumn{3}{c|}{\texttt{Eureka!} Aperture Photometry } & \multicolumn{3}{c}{\texttt{MIRIAM} Differential Photometry} \\
         
        \hline
        & $\mathbf{F_p\ (\textbf{ppm})}$ & $\mathbf{t_s\ (\textbf{min})}$ & \textbf{Res. RMS (ppm)} & $\mathbf{F_p\ (\textbf{ppm})}$ & $\mathbf{t_s\ (\textbf{min})}$ & \textbf{Res. RMS (ppm)} \\
        
        \textbf{E1}&  $307^{+55}_{-56}$  &  $-0.1^{+2.3}_{-2.8}$ & 778 & $295^{+56}_{-57}$ & $0.4^{+3.3}_{-3.9}$ & 795 \\

        \textbf{E2} &  $200^{+55}_{-56}$  &  $-4.6^{+2.7}_{-3.5}$ & 774 & $223^{+55}_{-55}$ & $-5.0^{+2.0}_{-3.2}$  & 793 \\

        \textbf{E3} &  $307^{+51}_{-51}$  &  $-3.9^{+3.7}_{-1.6}$ & 774 & $310^{+52}_{-53}$ & $-1.1^{+4.7}_{-1.6}$  & 798 \\

        \textbf{Joint} & $262^{+30}_{-30}$ & $-4.5^{+0.9}_{-0.9}$ & 776 & $271^{+31}_{-30}$ & $-4.1^{+0.8}_{-0.8}$ & 796  \\
        \hline
        \hline
    \end{tabular}
    \tablecomments{ $t_s$ is expressed in minutes since expected time of secondary eclipse. A negative value corresponds to an eclipse happening before the predicted time. The values given for the individual eclipses are from the preferred model.}
\end{table*}


\subsection{Results from \texttt{Eureka!}}
We present the detrended time series for our joint fit light curves in the middle panels of Figure~\ref{fig:raw-eureka}. The residuals are shown in the bottom panels.
We find an eclipse depth of 262$\pm$30\,ppm for the joint fit. The observed time of mid-transit for a zero-eccentricity orbit occurs 4.5$\pm$0.9 minutes before the predicted time. 

%
From the model comparison of eclipse 1  (Figure~\ref{fig:e1-compare-plot}), the best-fit model is the exponential (model \texttt{E}). We can see the detrending models that fit the exponential ramp at the beginning of the observations are favoured. The linear model is a very poor fit in comparison to the other models, hugely overestimating the eclipse depth and has a $\Delta$BIC of 270.56 compared to the best model. A single, identical detrending model applied to the three eclipses cannot effectively represent the systematics.
For eclipse 2 (Figure~\ref{fig:e2-compare-plot}), the best fit model is a double exponential (model $\texttt{DE}$). 
For eclipse 3  (Figure~\ref{fig:e3-compare-plot}), the best fit model is the linear (model \texttt{L}).
The model comparison plots show that not detrending in $x_0$ and $y_0$ yields lower BIC values and better fits for every eclipse. 
For the individual fits of  the light curves where the first and last groups of each integration were kept, we found that detrending in $x_0$ and $y_0$ was actually preferred in every case.



\subsection{Results from \texttt{MIRIAM}}
We present the detrended and jointly fitted light curves for the \texttt{MIRIAM} extractions in the middle panels of Figure~\ref{fig:raw-miriam}. The residuals are shown in the bottom panels.
The same best-fit models are preferred for the light curves extracted with \texttt{MIRIAM}, respectively \texttt{E}, \texttt{DE} and \texttt{L}, for the first, second and last eclipses. 
We find an eclipse depth of 271$^{+31}_{-30}$\,ppm for the joint fit. The observed time of mid-transit for a zero-eccentricity orbit occurs 4.1$^{+0.8}_{-0.8}$ minutes before the predicted time. 
%
%
Our reported eclipse depths for the joint fit from two independent pipelines are fully consistent, 262$_{-30}^{+30}$\,ppm for \texttt{Eureka!} and 271$_{-30}^{+31}$\,ppm for \texttt{MIRIAM}. The novel differential photometry method \texttt{MIRIAM} delivers independent confirmation that the eclipse depth of LHS 1140 c is consistent with a hot dayside and likely no atmospheric heat redistribution.
%
%

For the joint fits, we measure a point-to-point scatter of 777\,ppm and a residual RMS of 796\,ppm for the light curves from \texttt{MIRIAM}. Respectively, we measure 767\,ppm and 776\,ppm for the light curves from \texttt{Eureka!}. 
Those residuals are consistent with the photon-noise level of $\sim$864\,ppm predicted by the JWST Exposure Time Calculator, which also forecasts an eclipse-depth uncertainty of $\sim$30\,ppm after three visits, in agreement with the observations (Table \ref{tab:table1}). While speculative, this near photon-noise-limited performance is likely attributable to the quiescent nature of LHS~1140 \citep{cadieux2024}. 

We note that all joint eclipse measurements of LHS\,1140\,c reported in the literature to date (Figure \ref{fig:atm_models}, left panel)  are remarkably consistent, showing a standard deviation of only 10 ppm and a peak-to-peak variation of 30 ppm. This level of agreement testifies to the robustness of the various eclipse-extraction algorithms and provides a favourable outlook for forthcoming, shallower eclipse measurements of LHS\,1140\,b within the 500-hr \textit{Rocky Worlds} survey and other similar programs focused on relatively cold planets.



\subsection{Brightness Temperature}
%
%

To derive the brightness temperature, $T_B$, from the eclipse depth, we generate a series of  blackbody spectrum using Planck's law for a range of planet temperatures between 100\,K and 1000\,K. We compute the expected $F_p$ for each temperature value, then integrate over the MIRI/F1500W bandpass using the instrument throughput. We interpolate between $F_p$ and the planet temperature and use the posterior distribution of $F_p$ to calculate the corresponding $T_B$. We use the median value of the posterior distribution as the best fit value and compute the uncertainty from the 16$^{\rm th}$ and the 84$^{\rm th}$ percentiles of the distribution. 

We find a value of $T_B$ = 587$_{-33}^{+33}$\,K for the average value of the joint fit with \texttt{Eureka!} and of 595$_{-33}^{+34}$\,K for the fit with \texttt{MIRIAM}.
The expected temperature of LHS\,1140\,c for a zero-albedo blackbody with full heat redistribution is $T_B =425$\,K, while for a bare rock with no heat redistribution, \hbox{$T_B =543$\,K}. This is a simplified calculation that does not account for the Greenhouse effect or other atmospheric effects.
Our brightness temperature is consistent with that of \citet{fortune2025} (\hbox{$T_B =561\pm44$\,K}).

\section{Discussion}\label{sec:discussion}

\begin{figure*}
    \centering
    \includegraphics[width=1\linewidth]{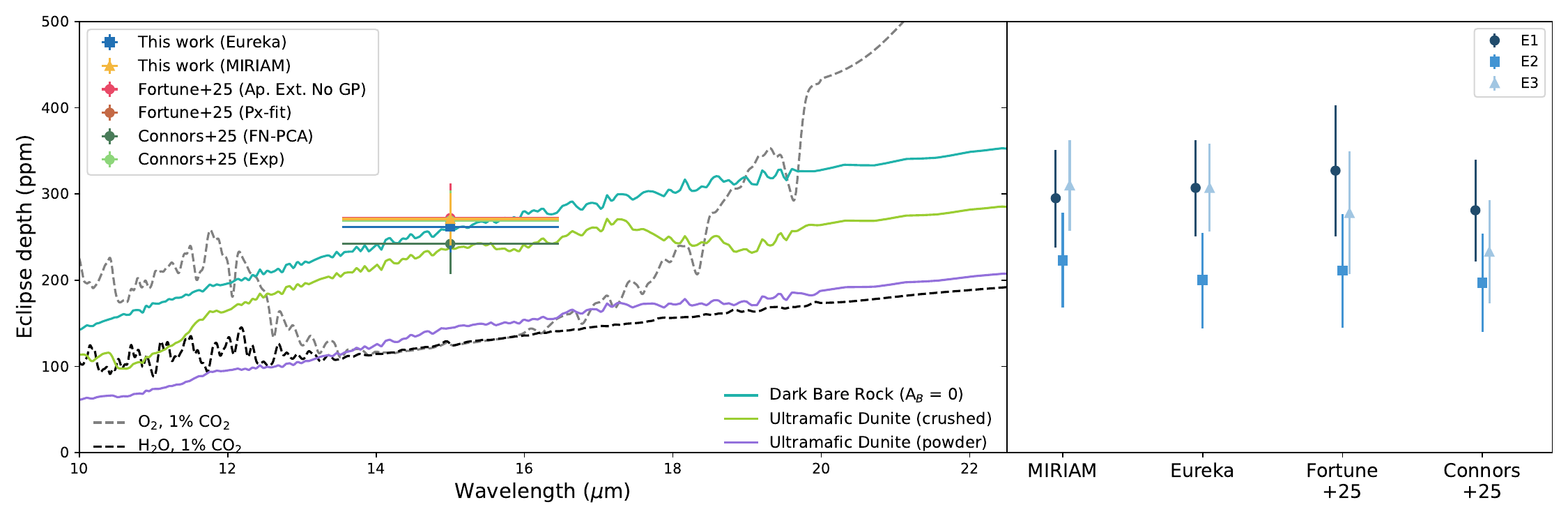}
    \caption{Left: Atmospheric models compared to the eclipse depth derived in this work and the values from the literature. The solid lines correspond to bare rock surface models and the dashed lines to atmospheric models.
    Right: The eclipse depth with errors for each eclipse derived in this work and from the literature. The data points are slightly offset horizontally for clarity. This figure was adapted from Figure 9 in \citet{connors2025}.}
    \label{fig:atm_models}
\end{figure*}

Figure~\ref{fig:atm_models} compares the measured eclipse depths from this work and from the literature. In the left panel, we plotted the joint eclipse depths against simulated emission spectra for different atmospheric compositions and planetary surfaces. The models are adapted from \citet{connors2025}. 
The bare rock models simulate the surface emissions of a bare rock exoplanet for a given surface composition and stellar model (Monaghan et al. (in prep.)). The Ultramafic Dunite spectra were modeled using reflectance measurements from \citet{Paragas2025}. The atmospheric models produce a nongray, radiative, convective temperature profile for a  well-mixed composition of 100\,ppm of CO$_2$, within an O$_2$ or H$_2$O dominated atmosphere, and compute the associated emission spectrum of the secondary eclipse \citep{benneke2012, benneke2013}.

The measured eclipse depths are all in agreement and are compatible with LHS\,1140\,c being a bare rock. The atmosphere scenarios are ruled out at $>$4$\sigma$.
This is also in agreement with the derived $T_B$.
In the right panel of Figure \ref{fig:atm_models}, we compare the individual value of $F_p$ for each eclipse. The measurements are consistent between all methods within 1$\sigma$ for the first two visits, and within 2$\sigma$ for the last visit.
The significant differences in the systematics seen in each individual eclipse adds a challenge for characterization. While multiple methods find agreeing eclipse depth, it is clear that multiple visits are required to constrain $F_p$. Although our analysis confirms the bare rock nature of LHS\,1140\,c, we can further investigate the implication of this measurement on the orbit of the planet and its internal structure when combined with insights from existing transit and stellar abundance measurements. 
%
%

%
\subsection{Eclipse Timing and Eccentricity}
We investigate the eclipse timing since the relative timing of the eclipse compared to the transit time can constrain the orbital eccentricity of the planet. For a zero-eccentricity fit, we find a time of mid-eclipse consistently a few minutes earlier than the expected $t_s$ for a circular orbit (see Table \ref{tab:table1}). 
Light travel time also is not accounted for by \texttt{batman}, representing a difference of $\sim$22\,s. 

We did a joint fit with the \texttt{MIRIAM} photometry, where $t_s$ was fixed to the predicted time for a circular orbit, which resulted in a lower BIC of $\Delta$BIC=2.4, with respect to the fit where $t_s$ is not fixed. This small $\Delta$BIC indicates that neither solution is preferred.
While the decrease in BIC is expected because the total number of free parameters is reduced, this is indicative that the fit is consistent with a value of $t_s$ close to the expected value for a circular orbit.
We use the following equation
\begin{equation}
    e\cos\omega \simeq \frac{\pi}{2P} (t_0 - t_s - \frac{P}{2}),
\end{equation}
to place an upper limit on the orbital eccentricity, $e$, of LHS\,1140\,c \citep{Charbonneau2005}. 
We do not fit for the eccentricity of the orbit, but we convert the time difference between the expected time of mid-eclipse from a circular orbit and the derived time of mid-eclipse, $t_s$, into an upper limit on the eccentricity. 
We derive a $3\sigma$ upper limit of $e$ $<$ 0.0019 for $t_s = -4.1\pm0.8$ for \texttt{MIRIAM} and $e < 0.0021$ for $t_s = -4.5\pm0.9$ for \texttt{Eureka!}. 

\subsection{Interior Structure Model}
Next, as the absence of an atmosphere simplifies interior structure inference, we aim to explore the possible interior composition of LHS\,1140\,c informed by the planet's known mass and radius and by measurements of the host stellar abundances. These methods are described in more detail in Weisserman et al. (in prep), and here we provide a brief summary.

\subsubsection{Inferring the Composition of LHS\,1140\,c from its Mass and Radius} \label{subsubsec:composition-mr}

We first perform an analysis of LHS\,1140\,c's composition solely informed by the planet's mass and radius. Based on our findings that the surface of LHS\, 1140\,c is consistent with a bare rock, we seek to constrain the planet's core-mass fraction (CMF). We model the planet's interior structure using the \texttt{exopie} package \citep{PlotnykovValencia2024}, which is an updated version of the earlier interior structure code \texttt{SuperEarth} \citep{PlotnykovValencia2020}. \texttt{exopie} assumes a differentiated, two-layered terrestrial planet composed of a primarily-iron core and a primarily-silicate mantle, although iron and silicon are free to partition into the mantle and core layers, respectively. We generate forward models of the planet given input mass, CMF, and $\texttt{xSi}$ and $\texttt{xFe}$ values representing the molar fractions of silicon in the core and iron in the mantle, respectively, which we allow to vary uniformly between $0-0.2$. We sample the joint $\{M_p, \texttt{CMF}, \texttt{xSi}, \texttt{xFe}\}$ prior space to find the planetary radius that best fits the observed radius. For this exercise, we adopt the mass and radius of LHS\,1140\,c to be $1.91 \pm 0.06 M_\oplus$ and $1.272 \pm 0.026 R_\oplus$, respectively, as reported by \citet{cadieux2024}. We sample the prior parameter space with $10^5$ draws to quantify the joint posterior, which includes the CMF parameter of interest. We find that the planet's CMF is consistent with $0$ assuming a two-layered interior and a 95\% upper limit on the CMF of $0.22$. 

It is clear that under our assumed model of a two-layered, core-mantle planetary interior, the planet's mass, radius, and absence of a thick atmosphere appear to imply an iron-depleted planet. However, given the observed connection between planetary and stellar refractory ratios \citep[e.g.,][]{PlotnykovValencia2020,AdibekyanDorn2021,AdibekyanDeal2024,BrinkmanPolanski2024}, this interpretation is unlikely to be correct unless LHS\,1140 is similarly depleted in iron relative to magnesium and silicon. We test this in the following subsection.

\subsubsection{Inferring the Composition of LHS\,1140\,c Informed by its Stellar Abundances} \label{subsubsec:composition-abund}

Here, we seek to model the CMF of LHS\,1140\,c using the measured refractory abundances of LHS\,1140.  Assuming that because the star and planet formed from the same primordial cloud, the planetary CMF should reflect the refractory abundance ratios of the host star, and so the stellar refractory abundances can be used as another method of characterizing its planets.

To perform this analysis, we use the refractory abundances of LHS\,1140 presented in Weisserman et al. (in prep). 
These stellar refractory abundances are measured from spectra obtained with the Near Infrared Planet Searcher \citep[NIRPS;][]{BouchyDoyon2025}.
For convenience, the refractory abundances for Fe and Ti are repeated in Table \ref{table:abundances-appendix}.

\begin{table}
\caption{Reported Stellar Abundances}
\centering
    \begin{tabular}{c c c c}
    \hline\hline
Abundance & \# Lines & [X/H] &  $\sigma_\textrm{[X/H]}$ \\
\hline
{[Fe/H]} & 17 & -0.250 & 0.110 \\
{[Ti/H]} & 35 & -0.304 & 0.120 \\
    \hline
    \end{tabular} \label{table:abundances-appendix}
    
    \smallskip
    \footnotesize

    This table details the results of the stellar abundance calculations discussed in Weisserman et al. (in prep).
\end{table}

We use \texttt{exopie}'s \texttt{star\_to\_planet()} function to sample forward models of LHS\,1140\,c's interior composition to solve for the planetary CMF whose refractory abundance ratios Fe/Mg and Mg/Si best match the observed stellar abundance ratios. We draw $10^4$ samples of the refractory abundances Mg, Si, and Fe from assumed Gaussian distributions set by the mean and standard deviation abundance values reported in Table \ref{table:abundances-appendix}. However, we note that we elect to replace the $\alpha$-element abundances [Mg/H] and [Si/H] with more reliable measurements of [Ti/H] (see Weisserman et al. (in prep) for details). The results of this calculation provide the posteriors of expected planetary CMF values, assuming that the stellar refractory abundances map directly to the interior of orbital terrestrial planets (i.e. with no other volatile species present).

Through this method, we find the expected CMF of LHS\,1140\,c  to be $0.34 \pm 0.11$. We note that this a value consistent with the Earth's CMF. This result seems to hold regardless of the stellar abundances used; \citet{cadieux2024} performs a separate analysis of LHS\,1140's refractory abundances, using the same NIRPS spectra as in Weisserman et al. (in prep), but using a different methodology. If we calculate the CMF of LHS\,1140\,c using the stellar abundances from this method, we find a CMF of $0.24 \pm 0.06$, a value consistent with the CMF found using the abundances from Weisserman et al. (in prep). Similarly, if instead we calculate the planetary CMF by adopting the mean abundance values from 31 M dwarfs in the solar neighborhood \citep{JahandarDoyon2025}, we measure a CMF of $0.25_{-0.12}^{+0.15}$. It is clear that the planetary CMFs predicted by different sources of stellar abundances yield consistent CMF values. What's more, these predicted CMF values are inconsistent with and systematically larger than the CMF measured from the planet's mass and radius. 

If these planets were purely made up of a rocky mantle and an iron core, this would suggest significant iron depletion of LHS\,1140\,c. However, planet formation models do not predict significant iron depletion of super-Earths \citep[e.g.,][]{ScoraValencia2020}. The refractory abundances of M dwarfs are largely homogeneous \citep{JahandarDoyon2025}, and the Mg, Si, and Fe that would make up the planet all have very similar condensation temperatures \citep{DornHarrison2019}, and so we would not expect these planets to have formed from iron-depleted material. As such, alternative explanations for these results are required. One potential explanation is that the planetary interior is contaminated with additional light elements that are not included in our present interior structure model, such as water. While this planet's equilibrium temperature is above the boiling point of water ($373\,\textrm{ K}$), recent models have suggested that water can be sequestered deep inside the interiors of super-Earths \citep{LuoDorn2024}. This can substantially inflate the planetary radius by up to $25\%$, in the process deflating the bulk density and thus the observed CMF from mass and radius, potentially explaining this discrepancy. As a similar water-rich composition has been proposed for LHS\,1140\,b as well \citep{cadieux2024,cadieux2024b,Damiano2024}, this seems an eminently plausible explanation for the composition of LHS\,1140\,c.


\subsection{Implication for Future Observations of the System}

It was announced that the system will be revisited with MIRI/imaging at 15\,$\mu$m under the Rocky Worlds DDT to observe secondary eclipses of LHS\,1140\,b. The measured timing of the eclipse is consistent with a nearly-circular orbit for LHS\,1140\,c, supporting the findings of \citet{fortune2025} and \citet{cadieux2024}. 
Theoretical predictions suggest that secular interactions with the inner planet LHS\,1140\,c should couple the eccentricities, rapidly damping them on a timescale $<1$\,Gyr, much shorter than the system's estimated age ($>5$\,Gyr; \cite{dittman2017}), with both planets likely tidally locked with their orbit circularized \citep{gomes2020}. However, the lack of a significant orbital eccentricity on LHS\,1140\,c, the inner planet, does not guarantee that the outer planet, LHS\,1140\,b, is also on a circular orbit. 
If LHS\,1140\,b had an orbital eccentricity equivalent to the derived upper limit for LHS\,1140\,c of $e<0.0019$, this would be equivalent to an eclipse timing uncertainty of 42.6\,min given the longer orbital period of the outer planet.
If we consider the upper limit on the eccentricity from \citet{fortune2025} of $e<0.0172$ at 95\% confidence obtained by fitting for $e\cos\omega$, we get an eclipse timing uncertainty of 390.0 min. Given the current empirical uncertainties on the orbital ephemerides of LHS\,1140\,b, it is still important to have a conservative baseline for the upcoming observations of LHS\,1140\,b as part of the DDT. 

The measured core-mass fraction of LHS\,1140\,c is compatible with the one reported for planet b in \citet{cadieux2024} of CMF = $20.5^{+5.5}_{-5.8}$ \% for stellar priors. It is expected that the planets will have agreeing CMF as they orbit the same star. This CMF is only compatible with the water world scenario for LHS\,1140\,b. 


To confirm the non-detection of the eclipse, it is essential to observe LHS\,1140\,b at another bandpass, insensitive to CO$_2$, like with the 21\,$\mu$m filter of MIRI (F2100W). This will confirm that the non-detection of the eclipse is due to the presence of an atmosphere and not a timing error from a potentially eccentric orbit of LHS\,1140\,b. Figure~5 of \citet{cadieux2024b} shows the predicted eclipse depth for LHS\,1140\,b given various atmospheric models and highlights the power of 21\,$\mu$m photometry to discriminate between the airless and CO$_2$-dominated case.
4 visits at 15\,$\mu$m will allow us to rule out the dark airless, scenario at \hbox{$\sim$4\,$\sigma$} \citep{cadieux2024b}. A H$_2$O-dominated atmosphere for LHS\,1140\,b has been excluded, so the observations will be immune from a muted H$_2$O feature like the one we see in the atmospheric model of LHS\,1140\,c in Figure~\ref{fig:atm_models} \citep{cadieux2024}.

\section{Conclusion}\label{sec:conclusion}
We re-analyzed three 15\,$\mu$m secondary eclipses of LHS\,1140\,c using two distinct methods of data reduction.
We presented a new data-driven framework to
extract light curves for JWST/MIRI photometric observations and compared the results to the classical method of aperture photometry.
We measured fully consistent eclipse depths of 262$_{-30}^{+30}$\,ppm for \texttt{Eureka!} and 271$_{-30}^{+31}$\,ppm for \texttt{MIRIAM}. This is in agreement with the already published values of the eclipse depth for LHS\,ll40\,c \citep{fortune2025, connors2025}.
We derived brightness temperatures of 587$_{-33}^{+33}$\,K for \texttt{Eureka!} and 595$_{-33}^{+34}$\,K for \texttt{MIRIAM}, compatible with an airless, rocky planet.
We find that the eclipse occurs 4.5$\pm$0.9 minutes before the expected time of mid-eclipse for \texttt{Eureka!} and 4.1$\pm$0.8 minutes before for \texttt{MIRIAM}. This is consistent with a nearly-circular orbit for LHS\,1140\,c, and in agreement with previous analyses and modeling \citep{cadieux2024, fortune2025, gomes2020}.
We find a CMF = 34$\pm0.11$\,\% from measured stellar abundances of refractory elements, inconsistent with the CMF of 0 from the mass and radius measurements. This could be explained by the presence of volatiles in the interior and would be in agreement with the suggested water-rich composition of LHS\,1140\,b.

The presence of significant systematics in the data that varies for each visit highlights the necessity of independent data reduction. Our fitting method of using an individual detrending model for each eclipse accounts for this variation. The novel approach with \texttt{MIRIAM} produces values in agreement with the literature and the commonly used method of aperture photometry.


For the upcoming observations of LHS\,1140\,b as part of the \textit{Rocky Worlds} DDT, we recommend a long baseline to account for a non-zero orbital eccentricity given that we find a small difference between the expected and measured time of eclipse of LHS\,1140\,c. We also recommend observing LHS\,1140\,b at a wavelength independent of CO$_2$ to confirm the detection of a flat eclipse, like 21$\mu$m.

\begin{acknowledgments}
A.R., L.D. and R.D. would like to acknowledge funding from the National Sciences and Research Council of Canada (NSERC), the Canadian Space Agency under contract 23JWGO2B06, and the Trottier Family Foundation to the Institut Trottier de recherche sur les exoplanètes. 
R.D. and \'EA acknowledge financial support from the Canadian Space Agency through grant number 22EXPJWST.
\end{acknowledgments}

\begin{contribution}

AR performed the data reduction, wrote the fitting algorithm and most of the manuscript.
\'EA contributed the \texttt{MIRIAM} algorithm and benchmarking performances against other data analysis tools.
DW performed the stellar abundance analysis and the internal structure model.
LD participated in the conceptualization of the project, provided initial code for occultation and systematic model fitting, residuals diagnostics and brightness temperature calculation, contributed to the paper edits and general advising.
RD participated in the conceptualization of the project, advised the development of the \texttt{MIRIAM} algortihm, contributed to the paper edits and general advising.
CC provided an independent eclipse analysis that yielded eclipse depths consistent with those reported in this paper.
RC contributed to the internal structure section.

\end{contribution}

%
\facilities{JWST(MIRI)}

\software{\texttt{astropy} \citep{astropy2013, astropy2018, astropy2022} \texttt{emcee} \citep{emcee}, \texttt{batman} \citep{Kreidberg2015}, \texttt{matplotlib} \citep{matplotlib}, \texttt{h5py} \citep{collette2013}, \texttt{Eureka!} \citep{Bell2022}, \texttt{numpy} \citep{Harris2020}, \texttt{scipy} \citep{Virtanen2020}, \texttt{MC3} \citep{cubillos2017}, \texttt{corner} \citep{corner}}

\bibliography{sample701,added_refs}{}
\bibliographystyle{aasjournalv7}


\newpage
\appendix
\counterwithin{figure}{section}

\section{The MIRI Analysis Module (\texttt{MIRIAM})}\label{sec:miriam-explanation}

The \texttt{MIRIAM} framework is inspired by the \texttt{SOSSISSE} \citep{lim2023} framework for time-resolved spectroscopy with NIRISS/SOSS \citep{doyon_niriss_2023}. The algorithm is based on the assumption that the point-spread function of the instrument (or the trace in the case of \texttt{SOSSISSE}) is stable throughout the time series, and that the astrophysical signal and systematics to be detrended are small perturbations that can be expressed as a linear expansion of derivatives with respect to the relevant quantities. The goal of \texttt{MIRIAM} is to provide a model that is as simple as possible while maintaining outputs that are directly traceable to physical quantities. We measure the differential photometry by describing the PSF's evolution through a Taylor expansion in terms of derivatives with respect to position and linearity. While not attempting to account for the underlying physical cause, this approach provides a simple model of spurious signals in MIRI TSO. The algorithm performs as well as significantly more complex algorithms while producing physically interpretable quantities.

The main steps of the algorithm are as follows:

\begin{itemize}
    \item We start with Stage~2 frames (rateints), but Stage~1 (uncal) data could also be used to allow for different methods of rate extraction (e.g., correlated double sampling instead of ramp fitting).
    \item MIRI images exhibit a common pattern of pixel offset every 4 pixels. We therefore subtract the median value of every fourth column of the full frame.
    \item We extract a region of interest (ROI) centered on the PSF for computation ($32\times32$\,pixels).
    \item A reference PSF is constructed by taking the median of the normalized frames. Normalization is performed by first computing a `raw' median (i.e., without prior normalization) of all frames, then finding the amplitude for each frame:
    \[
    a_{raw}(t) = \frac{\sum_{i,j} f_{i,j}(t) \cdot p_{i,j}}{\sum_{i,j} p_{i,j}^2},
    \]
    where $a_{raw}(t)$ is the normalization factor of the $n^{\rm th}$ frame, and $p$ is the `raw' point-spread function.
    \item The median PSF $M_{i,j}$ is used as a metric to measure the flux for frame $t$. We numerically compute the $x$ and $y$ derivatives of $M_{i,j}$. We further construct a shape function that describes the change in $M_{i,j}$ due to a non-linearity variation in the system. One could simply use $M^2_{i,j}$, but this would be strongly correlated with $M_{i,j}$ (since a squared PSF is morphologically similar to the PSF itself). We therefore define $D_{i,j} = M_{i,j}^2 - KM_{i,j}$, a constant PSF concentration residual, with $K$, a scalar chosen such that $D_{i,j}$ and $M_{i,j}$ are orthogonal (i.e., \hbox{$\sum_{i,j}M_{i,j}D_{i,j} = 0$} or, equivalently, $K = \sum_{i,j}M_{i,j}^3/\sum_{i,j}M_{i,j}^2$). By construction, if there is a small non-linearity change in the instrument between the observation of two PSFs, the residual of the least-squares fit of one PSF onto the other will have the structure of $D_{i,j}$ (i.e., $D_{i,j}$ multiplied by a constant). 
\end{itemize}

\begin{figure}
    \centering
    \includegraphics[width=0.5\linewidth]{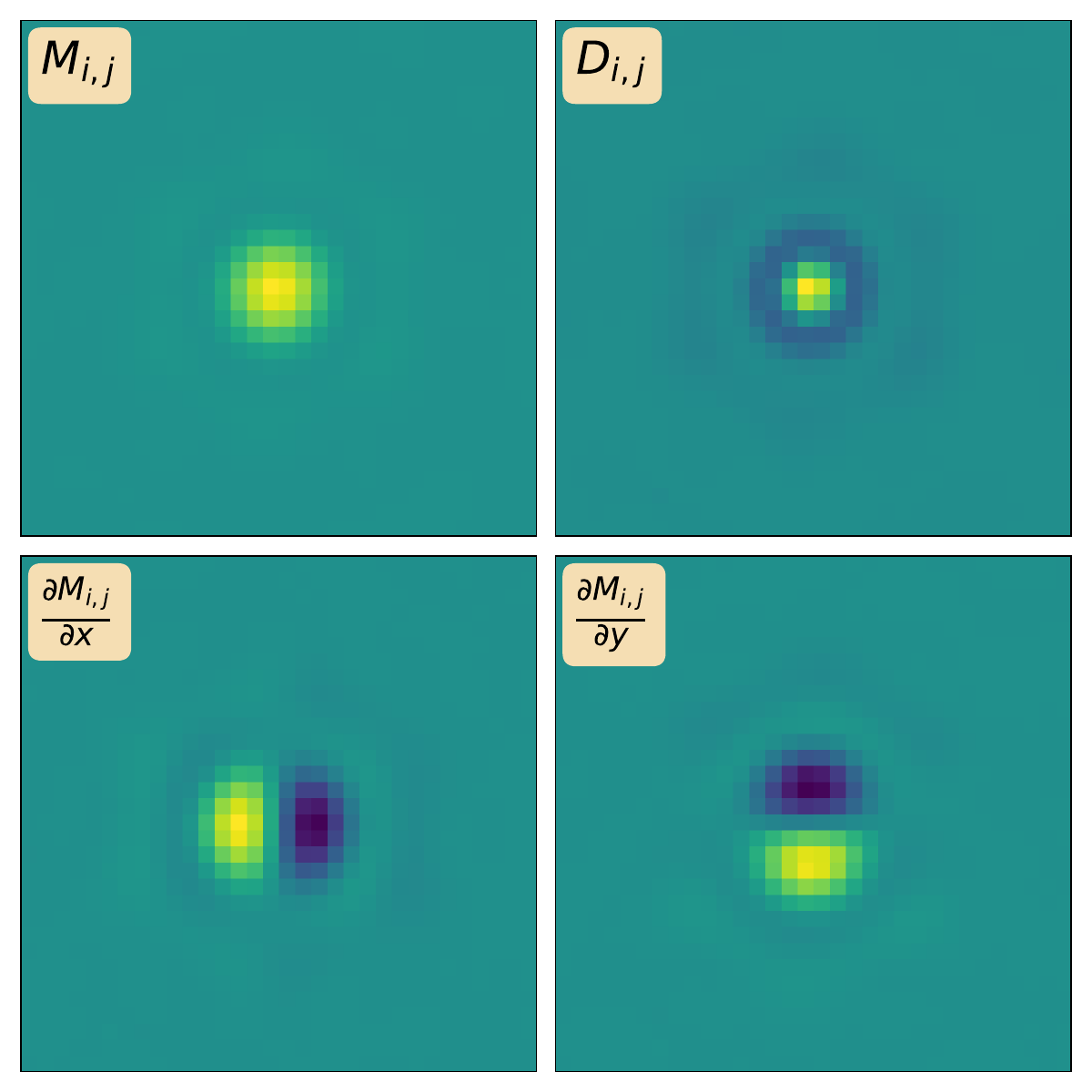}
    \caption{Point-spread-function and the associated terms in the \texttt{MIRIAM} framework for Eclipse~1. The data products are virtually identical for other eclipses. Color scaling is linear with flux.}
    \label{fig:miriampsf}
\end{figure}

\begin{figure}
    \centering
    \includegraphics[width=1.0\linewidth]{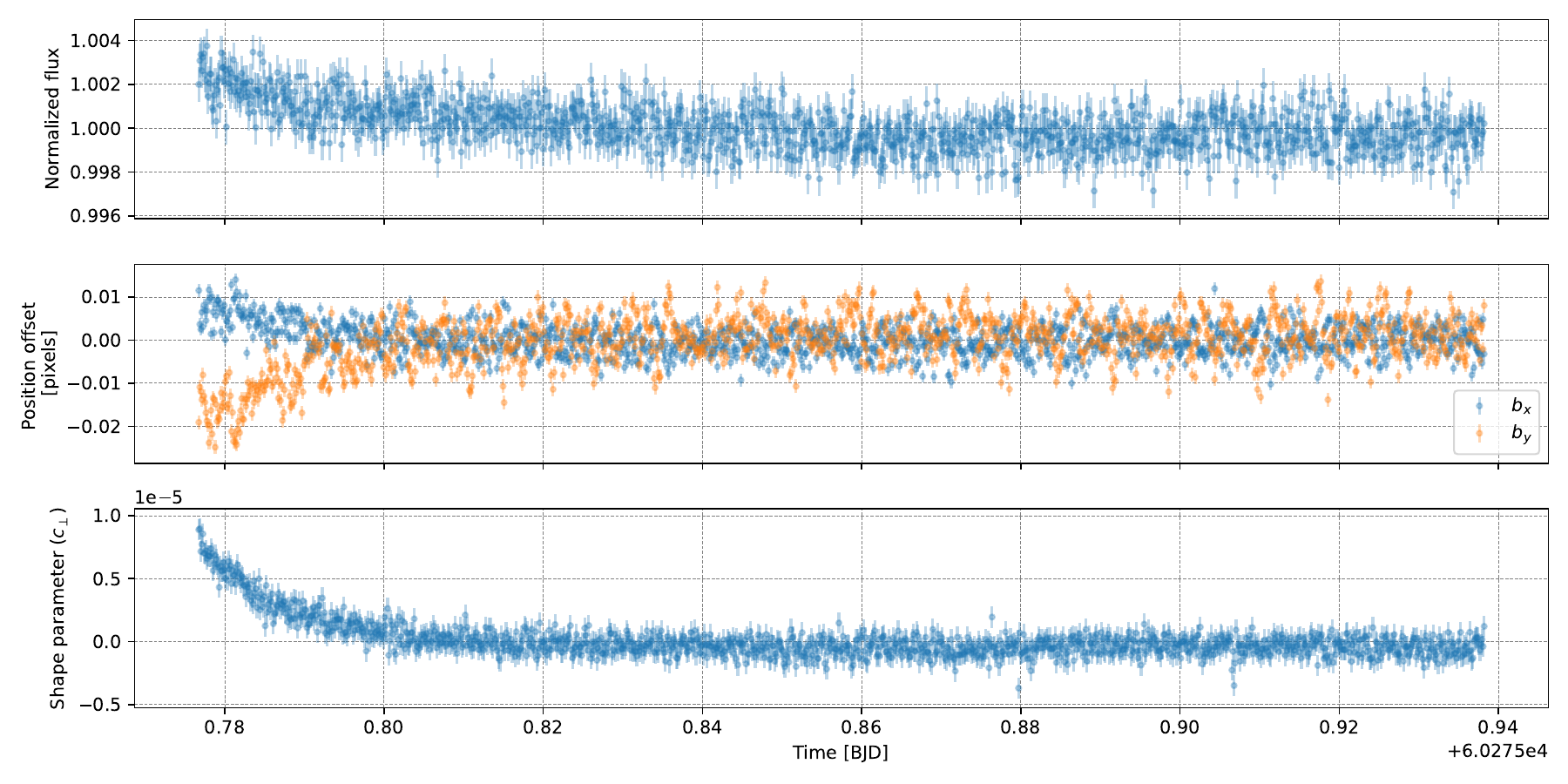}
    \caption{Time series of Eclipse~1 for detrending parameters from \texttt{MIRIAM}. Morphological changes in the PSF structure at the start of the sequence are readily seen in $c_\perp$, and to a lesser extent, in position.}
    \label{fig:miriam}
\end{figure}

The model image can be expressed as
\begin{equation}
I_{i,j}(t) = a(t) M_{i,j}  + b_x(t) \frac{\partial M_{i,j}}{\partial x} + b_y(t) \frac{\partial M_{i,j}}{\partial y} + c_\perp(t) D_{i,j},
\end{equation}

where $I_{i,j}(t)$ is the reconstructed image at time $t$ for pixels $i,j$, and $M_{i,j}$ is the template image constructed from all frames. Figure~\ref{fig:miriampsf} illustrates the morphology of the 4 terms used in the model image. By construction, $a(t)$ is an amplitude term that scales with the flux. The terms $b_x(t)$ and $b_y(t)$ directly measure the PSF's motion, while $c_\perp(t)$ traces morphological changes. All parameters can be used for detrending photometric measurements or, at the very least, to ensure that any correlation remains below a desired threshold. The model $I_{i,j}(t)$ is then used to measure the flux directly. Figure~\ref{fig:miriam} illustrates the time series of $b_x$, $b_y$, and $c_\perp$ for Eclipse~1.

\newpage
\section{Model Comparison}\label{sec:model_compare}
\begin{figure*}
\centering
\begin{subfigure}{1\textwidth}
    \includegraphics[width=\textwidth]{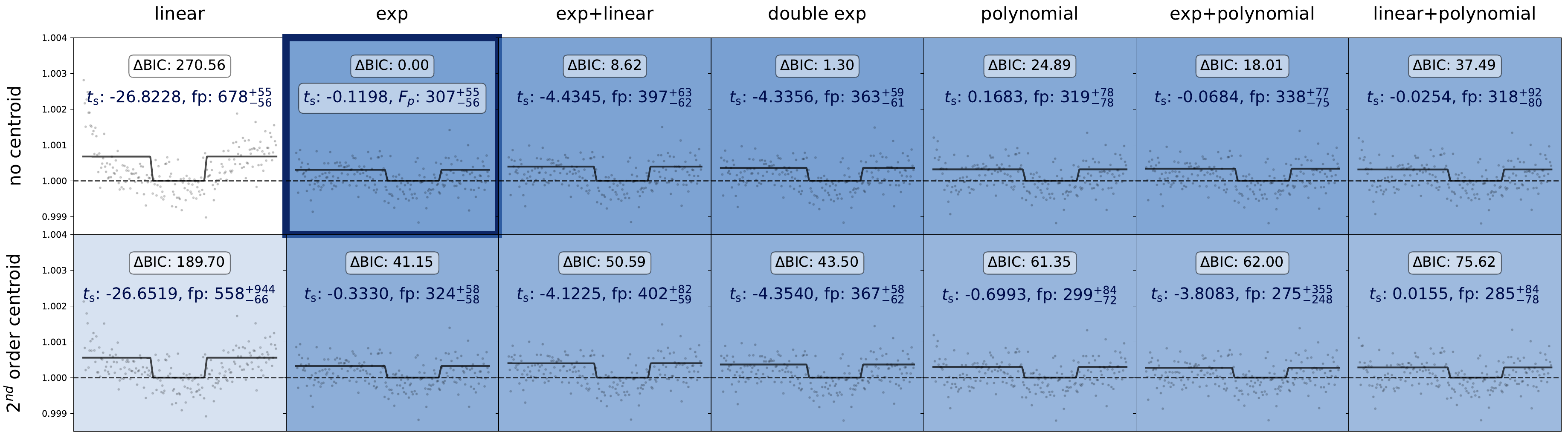}
    \caption{Model comparison for eclipse 1. The preferred fit uses an exponential function without a centroid model.} 
    \label{fig:e1-compare-plot}
\end{subfigure}
\hfill
\begin{subfigure}{1\textwidth}
    \includegraphics[width=\textwidth]{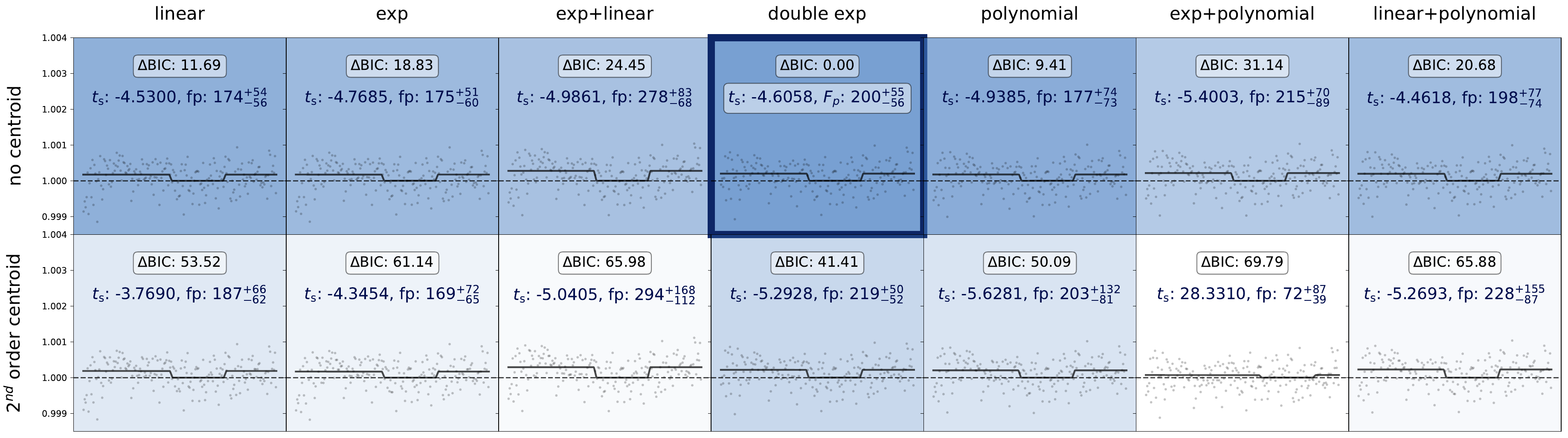}
    \caption{Comparative plot for eclipse 2. The preferred fit uses a double exponential function without a centroid model.}
    \label{fig:e2-compare-plot}
\end{subfigure}
\hfill
\begin{subfigure}{1\textwidth}
    \includegraphics[width=\textwidth]{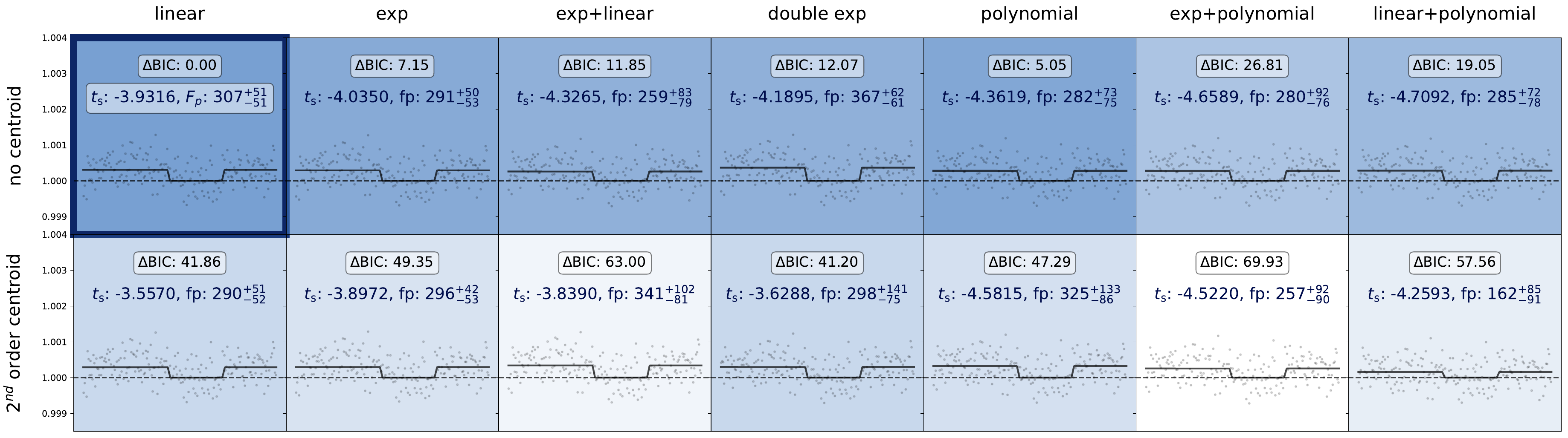}
    \caption{Comparative plot for eclipse 3. The preferred fit uses a linear function without a centroid model. }
    \label{fig:e3-compare-plot}
\end{subfigure}
        
\caption{Model comparison for the individual fits of each eclipse. The eclipse depth $F_p$ is expressed in ppm. The time of mid-eclipse, $t_s$, is expressed in minutes and negative values signify that the eclipse is occurring before the predicted time. The opacity of the box indicates the fit preference (darker is better). The preferred model is boxed in dark blue for each eclipse. The detector models as a function of time are shown as the columns, including the multiplied models. The rows indicate whether or not a detector model as a function of the centroid positions, $x_0$ and $y_0$ was used.}
\label{fig:compare_plots}
\end{figure*}

The comparative grid to show all models tested for each eclipse is shown in Figure~\ref{fig:compare_plots}. Each column corresponds to one of the different detrending model as a function of time described in section \ref{sec:detrending_models}. The rows compare the use of the second-order polynomial centroid model. The preferred model (lowest BIC) for each eclipse is boxed in dark blue.

\section{Red-Noise Tests}
\begin{figure*}
\centering
\begin{subfigure}{1\textwidth}
    \includegraphics[width=\textwidth]{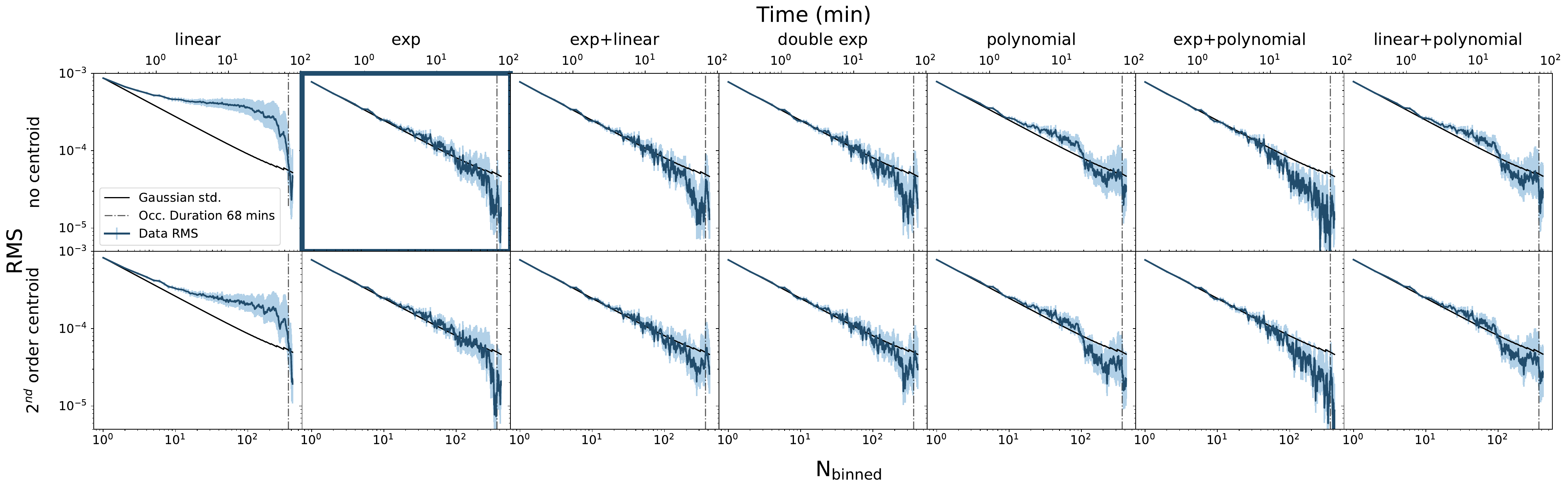}
    \caption{Eclipse 1}
    \label{fig:first}
\end{subfigure}
\hfill
\begin{subfigure}{1\textwidth}
    \includegraphics[width=\textwidth]{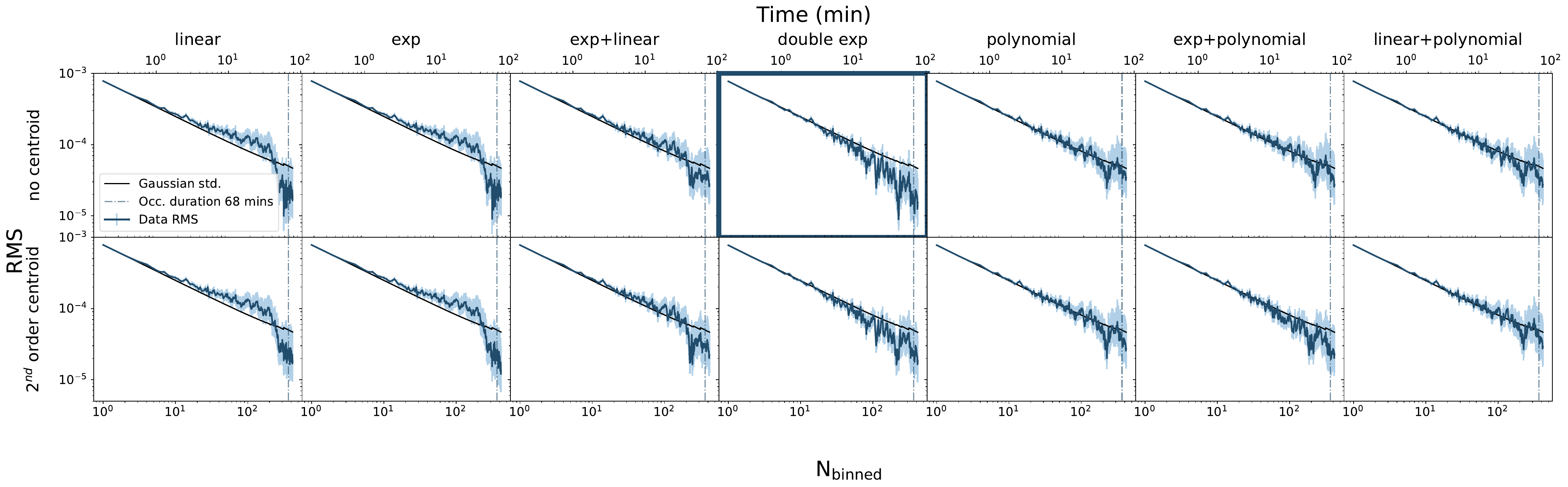}
    \caption{Eclipse 2}
    \label{fig:second}
\end{subfigure}
\hfill
\begin{subfigure}{1\textwidth}
    \includegraphics[width=\textwidth]{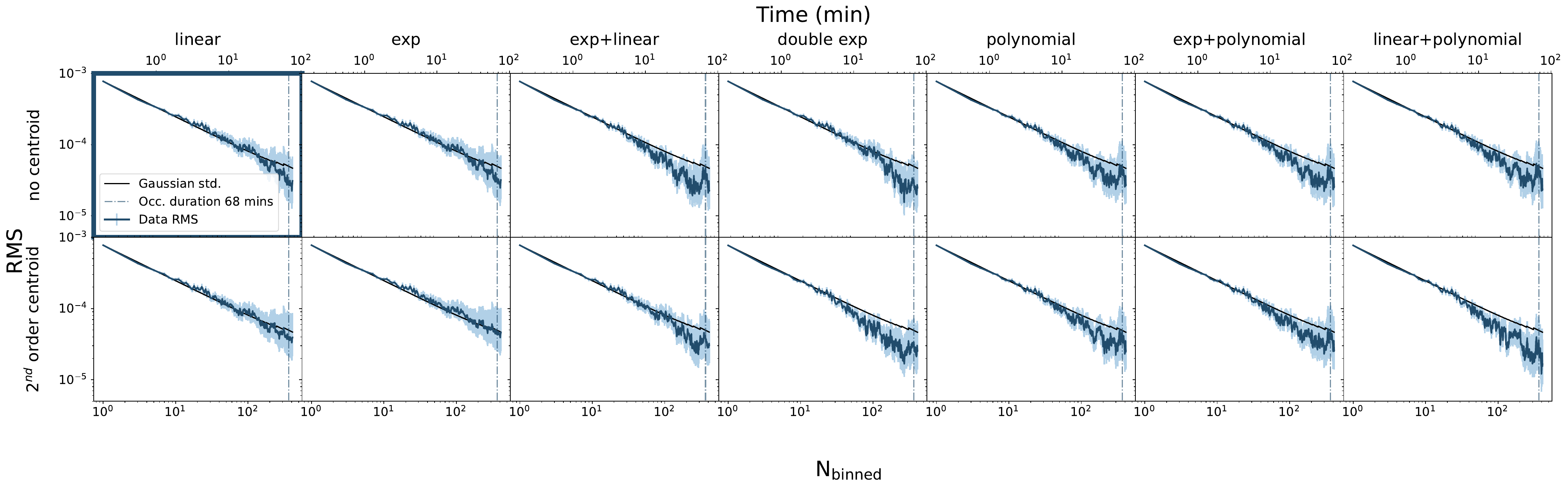}
    \caption{Eclipse 3}
    \label{fig:third}
\end{subfigure}
        
\caption{Comparative red-noise test for each model as a function of bin-size for the three eclipses with the \texttt{Eureka!} photometry. We binned the data at increasing bin sizes, $N_{\text{binned}}$ and calculated the RMS of the binned residuals. The light blue shaded region corresponds to the uncertainty (computed using the MC3 package \citep{cubillos2017}). The vertical dashed line represents the number of bins contained in the duration of an eclipse depth. The black line is the expected decrease for purely white noise.}
\label{fig:rednoise-comparison}
\end{figure*}

\begin{figure*}
\centering
\begin{subfigure}[b]{0.48\textwidth}
    \includegraphics[width=\textwidth]{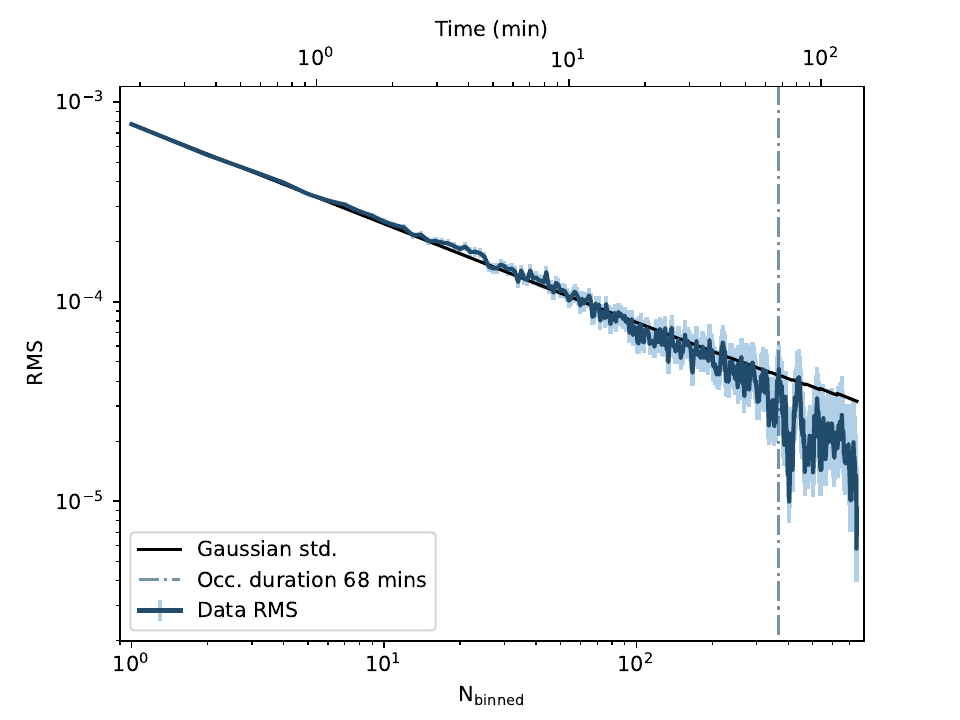}
    \caption{Red-noise test for joint fit with \texttt{Eureka!} photometry.}
    \label{fig:rednoise-eureka}
\end{subfigure}
\hfill
\begin{subfigure}[b]{0.48\textwidth}
    \includegraphics[width=\textwidth]{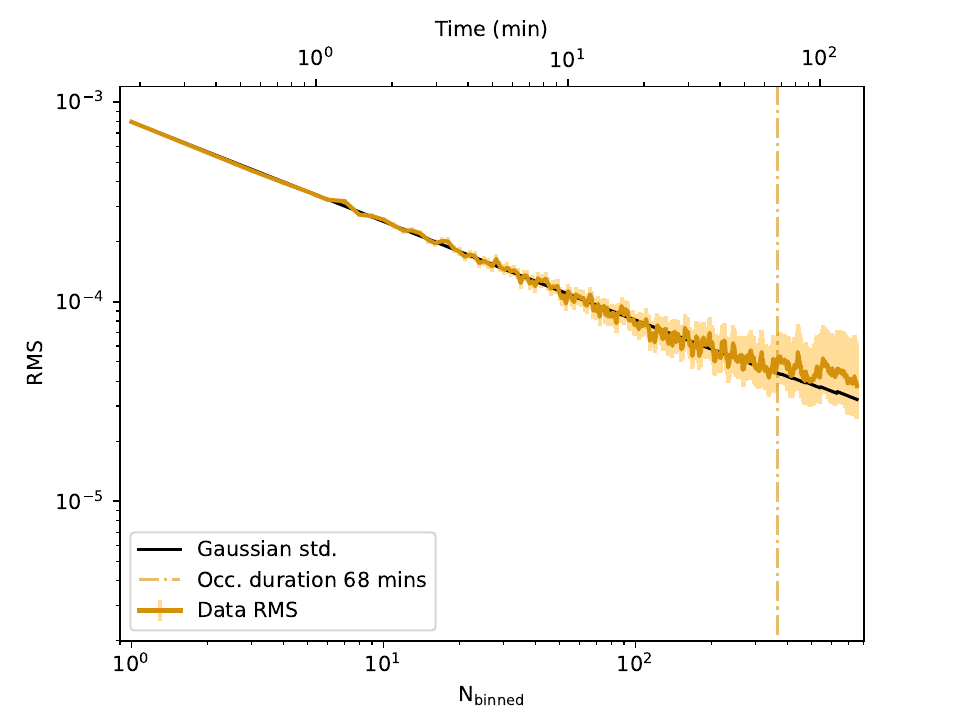}
    \caption{Red-noise test for joint fit with \texttt{MIRIAM} photometry.}
    \label{fig:rednoise-miriam}
\end{subfigure}
\caption{Comparison of the red-noise tests for the joint fit with both data reduction methods.}\label{fig:rednoise}
\end{figure*}

We present in Figure~\ref{fig:rednoise-comparison} the comparative red-noise tests that identify models with residual signals not captured by the fit.
The black line in each panel represents the expected decrease in residual RMS for pure white noise. The vertical dashed line corresponds to the eclipse duration. We also show the red-noise test for both joint fits, with \texttt{Eureka!} and \texttt{MIRIAM} in Figure~\ref{fig:rednoise}. While both methods are in good agreement with the expected decrease in RMS, the \texttt{MIRIAM} joint fit approximates the black line more closely, indicative that the astrophysical and detector variations are better fitted.  

\newpage

\end{document}